\documentclass[12pt,twoside]{article}
\usepackage[T1]{fontenc}
\usepackage[latin1]{inputenc}
\usepackage{times}
\usepackage{graphicx}
\usepackage{a4wide}
\usepackage{listings}
 \usepackage{amsmath}

\begin{document}


\section*{\center Charged pion production in C+C and Ar+KCl collisions \\ 
measured with HADES
\footnote{Contribution presented at the XLVII International 
Winter Meeting on Nuclear Physics, Bormio (Italy), Jan. 26-30, 2009}}

\thispagestyle{empty}

\markboth{P.~Tlust\'{y} {\it et al.}}
{Charged pion production in C+C and Ar+KCl collisions measured 
with HADES}

P.~Tlust\'{y}$^{14}$, G.~Agakishiev$^{8}$,
A.~Balanda$^{3,e}$, G.~Bellia$^{1,a}$, D.~Belver$^{15}$, A.~Belyaev$^{6}$,
A.~Blanco$^{2}$, M.~B\"{o}hmer$^{11}$, J.~L.~Boyard$^{13}$,
P.~Braun-Munzinger$^{4}$
% ...

\begin{center} (HADES collaboration)
\end{center}

\begin{raggedright}

\hspace{-0.4cm}\makebox[0.3cm][r]{$^{1}$}
Istituto Nazionale di Fisica Nucleare - Laboratori Nazionali del Sud, 95125~Catania, Italy\\
\hspace{-0.4cm}\makebox[0.3cm][r]{$^{2}$}
LIP-Laborat\'{o}rio de Instrumenta\c{c}\~{a}o e F\'{\i}sica Experimental de Part\'{\i}culas, 3004-516~Coimbra, Portugal\\
\hspace{-0.4cm}\makebox[0.3cm][r]{$^{3}$}
Smoluchowski Institute of Physics, Jagiellonian University of Cracow, 30-059~Krak\'{o}w, Poland\\
\hspace{-0.4cm}\makebox[0.3cm][r]{$^{4}$}
Gesellschaft f\"{u}r Schwerionenforschung mbH, 64291~Darmstadt, Germany\\
\hspace{-0.4cm}\makebox[0.3cm][r]{$^{5}$}
Institut f\"{u}r Strahlenphysik, Forschungszentrum Dresden-Rossendorf, 01314~Dresden, Germany\\
\hspace{-0.4cm}\makebox[0.3cm][r]{$^{6}$}
Joint Institute of Nuclear Research, 141980~Dubna, Russia\\
\hspace{-0.4cm}\makebox[0.3cm][r]{$^{7}$}
Institut f\"{u}r Kernphysik, Johann Wolfgang Goethe-Universit\"{a}t, 60438 ~Frankfurt, Germany\\
\hspace{-0.4cm}\makebox[0.3cm][r]{$^{8}$}
II.Physikalisches Institut, Justus Liebig Universit\"{a}t Giessen, 35392~Giessen, Germany\\
\hspace{-0.4cm}\makebox[0.3cm][r]{$^{9}$}
Istituto Nazionale di Fisica Nucleare, Sezione di Milano, 20133~Milano, Italy\\
\hspace{-0.4cm}\makebox[0.3cm][r]{$^{10}$}
Institute for Nuclear Research, Russian Academy of Science, 117312~Moscow, Russia\\
\hspace{-0.4cm}\makebox[0.3cm][r]{$^{11}$}
Physik Department E12, Technische Universit\"{a}t M\"{u}nchen, 85748~M\"{u}nchen, Germany\\
\hspace{-0.4cm}\makebox[0.3cm][r]{$^{12}$}
Department of Physics, University of Cyprus, 1678~Nicosia, Cyprus\\
\hspace{-0.4cm}\makebox[0.3cm][r]{$^{13}$}
Institut de Physique Nucl\'{e}aire d'Orsay, CNRS/IN2P3, 91406~Orsay Cedex, France\\
\hspace{-0.4cm}\makebox[0.3cm][r]{$^{14}$}
Nuclear Physics Institute, Academy of Sciences of Czech Republic, 25068~Rez, Czech Republic\\
\hspace{-0.4cm}\makebox[0.3cm][r]{$^{15}$}
Departamento de F\'{\i}sica de Part\'{\i}culas, University of Santiago de Compostela, 15782~Santiago de Compostela, Spain\\
\hspace{-0.4cm}\makebox[0.3cm][r]{$^{16}$}
Instituto de F\'{\i}sica Corpuscular, Universidad de Valencia-CSIC, 46971~Valencia, Spain\\ 
\hspace{-0.4cm}\makebox[0.3cm][r]{$^{a}$}
{Also at Dipartimento di Fisica e Astronomia, Universit\`{a} di Catania, 95125~Catania, Italy}\\
\hspace{-0.4cm}\makebox[0.3cm][r]{$^{b}$}
{Also at ISEC Coimbra, ~Coimbra, Portugal}\\
\hspace{-0.4cm}\makebox[0.3cm][r]{$^{c}$}
{Also at Technische Universit\"{a}t Dresden, 01062~Dresden, Germany}\\
\hspace{-0.4cm}\makebox[0.3cm][r]{$^{d}$}
{Also at Dipartimento di Fisica, Universit\`{a} di Milano, 20133~Milano, Italy}\\
\hspace{-0.4cm}\makebox[0.3cm][r]{$^{e}$}
{Also at Panstwowa Wyzsza Szkola Zawodowa, 33-300~Nowy Sacz, Poland}

\end{raggedright}

\vspace{1.0cm}

\begin{center}
\textbf{Abstract}
\end{center}
Results of a study of charged pion production in 
$^{12}$C~+~$^{12}$C collisions at incident beam energies of 1A~GeV
and 2A~GeV, and $^{40}$Ar~+~$^{nat}$KCl at 1.76A~GeV, using the 
spectrometer HADES at GSI, are presented. 
We have performed a measurement of the transverse momentum distributions
of $\pi^\pm$ mesons covering a fairly large rapidity interval, in case of 
the C+C collision system for the first time.
The yields, transverse mass and
angular distributions are compared with
a transport model as well as with existing data from
other experiments.


\section{Introduction}

The study of particle production in nucleus-nucleus
collisions at relativistic energies is essential
for understanding the dynamics and the
approach of the system towards equilibrium 
and the generation of flow phenomena,
as well as for gaining information about the nuclear equation of state.
In heavy-ion collisions, pion spectra
and yields are affected by collective effects like thermalization,
directed and elliptic flow, as well as by possible modifications of the
properties of baryon resonances they stem from, in particular
the $\Delta$.
The subtle interplay of these phenomena is indeed a challenge to 
theoretical interpretations.

Best suited for a description of all phases of the complex 
dynamics of heavy-ion reactions
are transport models, based on microscopic kinetic theory.
Transport models achieve a remarkable success in describing bulk properties
of the interactions over a large energy and system size scale.
At the same time, however, for special channels, problems are met in reproducing
precisely the experimental data.
For a recent comprehensive discussion of various
differential pion observables and their comparison with model calculations
in the region of 1A~GeV see \cite{reisdorf}.

The~~High~~Acceptance~~ Di-Electron~~Spectrometer (HADES) \cite{ref2}, is
designed for high-resolution and high-acceptance dielectron
spectroscopy in hadron-hadron, ha\-dron-nucleus, and nucleus-nucleus
reactions at beam energies in the range from 1A~GeV to 2A~GeV.
For a detailed description of the device and recently performed 
experiments see 
\cite{bormio2009_froehlich}.

In this paper, we discuss data on charged pions
obtained from $^{12}$C~+~$^{12}$C collisions at 1A~GeV and 2A~GeV 
(recently published in \cite{CCpions}) 
and $^{40}$Ar~+~$^{nat}$KCl at 1.76A~GeV.  
In case of the $^{12}$C~+~$^{12}$C system, large intervals of rapidity 
and of centre-of-mass angle are covered 
for the first time.
Our results are compared to UrQMD transport-model predictions and 
experimental data from other experiments.

\section{Experiment}

The presented data were collected under the LVL1 trigger condition which was 
based on a fast determination of the
charged-particle multiplicity ($M_{ch}$) in the time-of-flight detectors.
The condition $M_{ch} \ge 4$ for $^{12}$C~+~$^{12}$C and $M_{ch} \ge 16$ 
for $^{40}$Ar~+~$^{nat}$KCl was used. 

In the first experiment, the collision system $^{12}$C~+~$^{12}$C at 2A~GeV was
studied with a beam intensity of $I_{beam} = 10^6$ particles/sec
impinging on a two-fold segmented carbon target with thickness of 
$2 \cdot 2.5\%$
interaction length.
In the second data taking period the $^{12}$C~+~$^{12}$C system was studied 
at 1A~GeV.
Then, for the first time, a high-resolution tracking mode exploiting
also the outer MDC planes was available. In this measurement, a carbon beam of
$10^6$ particles/sec was focused onto a carbon foil of 3.8\% interaction length.
In the last experiment analyzed here, the four-fold segmented target was made 
of natural KCl and has a total interaction length of 3.05\%. The $^{40}$Ar beam
intensity was about $6 \cdot 10^5$ particles/sec.

Particle identification in the HADES data analysis (for details see
\cite{ref2}) is based on Bayesian statistics. The method
allows to evaluate the probability that the reconstructed track can
be related to a certain particle species (e.g. proton, kaon, pion, 
electron, etc.). It combines several observables from various
sub-detectors (e.g. time-of-flight, energy loss) via probability
density functions determined for each observable and for all
possible particle species. 
In our case, hadron identification has been performed
using measured momenta and corresponding velocities computed by means of the
time-of-flight. 
After the particle identification is done for all tracks, the
resulting yields are corrected for efficiency and purity of the PID
method, as well as for the detector and tracking efficiencies.
The detection/tracking efficiency has been obtained from Monte
Carlo simulated and reconstructed UrQMD events.
Further experimental details can be found in \cite{CCpions,krizek}.

\begin{center}
 \begin{figure*}[htb!]
  \vspace*{+.01cm}
  \includegraphics*[width=130mm]{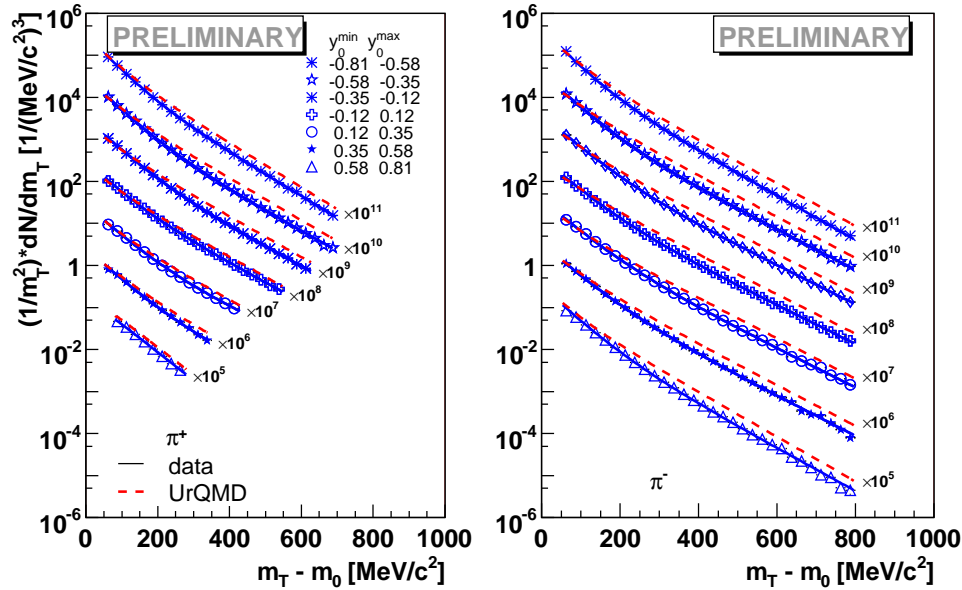}
  \vspace*{-0.1cm}
  \caption[]{Transverse-mass distributions for positively (left) and negatively
  (right) charged pions in different slices of rapidity derived from the data
  in $^{40}$Ar~+~$^{nat}$KCl at 1.76A~GeV (LVL1 ``semicentral'' events).
  Full lines show the results of fits to the data (symbols) using two 
  exponential 
  functions, while dashed lines show fits of the UrQMD distributions using 
  the same function. Error bars (systematic and statistical ones) are not
  visible at this scale. Both data and UrQMD distributions are normalized
  to the number of LVL1 events.

}
  \label{dNdMpion}
\end{figure*}
\end{center}

\section{Results}

\subsection{Transverse-mass distributions}

Figure~\ref{dNdMpion} exhibits the measured and simulated
transverse mass distributions of $\pi^{+}$ and $\pi^{-}$ in different 
intervals of normalized rapidity $y_0$
for the reaction $^{40}$Ar~+~$^{nat}$KCl at 1.76A~GeV.

The transverse mass distributions for all three data sets can be described
by a single-exponential (${}^{12}$C~+~${}^{12}$C at 1A~GeV) or 
two-exponential (${}^{12}$C~+~${}^{12}$C at 2A~GeV 
and $^{40}$Ar~+~$^{nat}$KCl) functions.

As shown in \cite{CCpions}, our data for ${}^{12}$C~+~${}^{12}$C
reactions are in agreement 
with the data measured by KaoS \cite{kaos}. For $^{40}$Ar~+~$^{nat}$KCl
at 1.76A~GeV (see 
fig.~\ref{dNdMpion}) our data also agree with results from 
the Ar+KCl at 1.8A~GeV reaction \cite{brockman}.
As seen in fig.~\ref{dNdMpion}, the agreement of UrQMD simulations with 
our data is 
only qualitative. Both data sets for $\pi^{+}$ and $\pi^{-}$ are described 
by two-exponential shapes.
However, the UrQMD simulations show a much 
stronger stiff component at large momenta.
In case of ${}^{12}$C~+~${}^{12}$C at 1A~GeV the agreement is much better, 
and both UrQMD and data 
exhibit a single-exponential shape. 
For ${}^{12}$C~+~${}^{12}$C at 2A~GeV, UrQMD predicts different spectral 
shapes -
purely exponential as compared to the concave shape of the data
\cite{CCpions}.
Data of $\pi^{0}$ for ${}^{12}$C~+~${}^{12}$C at 2A~GeV also need only  a 
single-exponential fit \cite{taps}.
This may indicate differences in the reaction dynamics 
of charged and neutral $\pi$ mesons, not described properly in 
the employed transport code.

\subsection{Multiplicities}

Pion yields ($N_{\pi}$) per reaction (under the LVL1 trigger condition) within
the HADES acceptance region and extrapolated to
 full phase space are presented in Table~\ref{tab1}.
The last column shows the multiplicities per participant, obtained by dividing
the yields with the averaged number of 
participating nucleons in the LVL1-triggered reactions estimated from 
simulations.

\begin{table}[h!]
\caption{
Production yields of $\pi^\pm$ per reaction under LVL1 trigger 
condition. N$_\pi$ and N$_\pi$(4$\pi$) are the measured and 4$\pi$ 
extrapolated yields, respectively, and $N_{\pi}(4\pi)/N_{part}$ is the 
yield per participant averaged for $\pi^+$ and $\pi^-$. 
The statistical errors are negligible. Shown are the systematic errors 
due to the various efficiency corrections and the 4$\pi$ extrapolation 
(see text).
\label{tab1}}
\center~
\begin{tabular}{c c c c c c}
\hline
system &  energy & &  $N_{\pi}$ & $N_{\pi}(4\pi$) & $N_{\pi}(4\pi)/N_{part}$\\
 & (A GeV) &  &  &  & $(\cdot 10^{3})$ \\
\hline
${}^{12}$C~+~${}^{12}$C & 1 & $\pi^+$ & 0.36$\pm$0.02 & 0.46$\pm$0.03$\pm$0.05&55$\pm$3$\pm$5$\pm$4 \\
    &  & $\pi^-$ & 0.38$\pm$0.02 & 0.49$\pm$0.03$\pm$0.05& \\
\hline
${}^{12}$C~+~${}^{12}$C & 2 & $\pi^+$ & 0.77$\pm$0.04  & 1.19$\pm$0.06$\pm$0.11&147$\pm$7$\pm$13$~^{+0}_{-21}$ \\
    &  & $\pi^-$ & 0.82$\pm$0.04  & 1.28$\pm$0.06$\pm$0.12& \\
\hline
$^{40}$Ar~+~$^{nat}$KCl & 1.76 & $\pi^+$ & 1.67$\pm$0.08  & 3.00$\pm$0.15$\pm$0.27&89$\pm$4$\pm$8$\pm$6\\
       &     & $\pi^-$ & 2.08$\pm$0.10  & 3.84$\pm$0.19$\pm$0.34& \\
\hline
\end{tabular}\\[3pt]
\end{table}

The systematic error of the measured yield due to uncertainties 
in the detection/reconstruction/identification  efficiency is estimated 
as $5\%$, based on a comparison of measurements in the six independent 
HADES sectors.
The extrapolation of the yields to full phase space is based on the
integration of the simulated rapidity distributions 
and normalized to the data in the rapidity range covered
by HADES. 
Varying the input conditions of the simulations, the differences between 
the rapidity distributions 
give us an estimate of the systematic error of the yield extrapolations of 
$9\%$.
The last systematic error of the resulting $\pi^\pm$ multiplicity per 
participant (last column in Table ~\ref{tab1})
stems from  uncertainties connected with the determination of
the average number of participating nucleons using UrQMD simulation (7\%).

\subsection{Angular distributions}

\begin{center}
 \begin{figure*}[htb!]
{
  \vspace*{+.1cm}
  \includegraphics*[width=130mm]{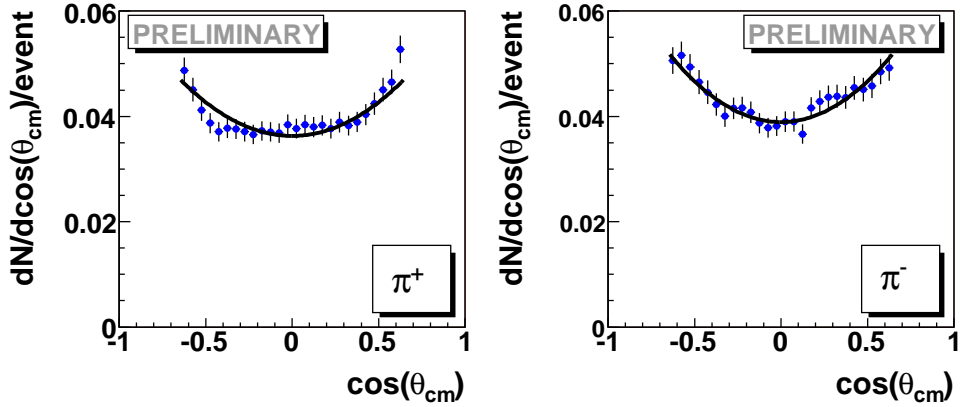}
  \vspace*{-0.1cm}
}
 \caption[]{Polar angle distribution in the center-of-mass system of
   positively (left) and negatively (right) charged $\pi$ mesons produced in
   $^{40}$Ar + $^{nat}$KCl collisions at 1.76A~GeV
   for the LVL1 (semicentral) events.
   Pions with center-of-mass momenta 200~-~800~MeV/c have been selected.
   The full lines show the fit as described in the text.
}
  \label{pid4cm}
\end{figure*}
\end{center}

The measured centre-of-mass polar angular distributions of pions
produced in $^{40}$Ar + $^{nat}$KCl collisions at 1.76A~GeV 
are exhibited in fig.~\ref{pid4cm}.
For the symmetric collision system the polar 
distributions in the
center-of-mass system can be fitted with the expression
$dN/d(\cos\theta_{cms})  \propto (1 + A_2 \cos^2 \theta_{cms} )$.
The  parameter $A_2$ characterizes the anisotropy of the pion
source. As visible in fig.~\ref{pid4cm},
the data show strong anisotropies quantified by 
$A_2 = 0.75 \pm 0.11$. Similar results are observed in the 
${}^{12}$C~+~${}^{12}$C system,
where $A_2 = 0.88 \pm 0.12$ and $1.19 \pm 0.16$
for beam energies of 1A~GeV and 2A~GeV, respectively. 
The two closest systems
studied most comprehensively in this respect are Ar+KCl at
1.8A~GeV \cite{brockman} and Ca+Ca at 1.93A~GeV \cite{reisdorf}.
In both cases similar momen\-tum-averaged anisotropies were observed, 
with values of $\langle A_2 \rangle = 0.5 - 0.6$.
Also the UrQMD simulations exhibit similar results for all three systems.

\section{Summary}

In summary, the charged-pion characteristics in the reactions
${}^{12}$C~+~${}^{12}$C at 1A~GeV and 2A~GeV, and $^{40}$Ar + $^{nat}$KCl 
at 1.76A~GeV have been measured in detail with the HADES
spectrometer. 
In case of ${}^{12}$C~+~${}^{12}$C reactions our data were obtained 
for the first time in a large acceptance region, 
which allowed for the measurement 
of pion anisotropy and consequently for a more reliable extrapolation 
of the yields to full solid angle. 
The found results are in good agreement with data obtained with 
previous experiments.
A comparison with the results on neutral pions and the UrQMD predictions for
${}^{12}$C~+~${}^{12}$C at 2A~GeV suggests differences in the reaction 
dynamics of charged and neutral $\pi$ mesons, not yet described by transport 
codes.
 
\section*{Acknowledgments} 

The HADES collaboration gratefully
acknowledges the support by BMBF grants 06MT238, 06TM970I, 06GI146I, 
06F-140, 06FY171, and 06DR135, 
by DFG EClust 153 (Germany), 
by GSI (TM-KRUE, TM-FR1, GI/ME3, and OF/STR), 
by grants GA AS CR IAA100480803 and MSMT LC 07050 (Czech Republic), 
by grant KBN 5P03B 140 20 (Poland), 
by INFN (Italy), by CNRS/IN2P3 (France), by 
grants~~ MCYT~~ FPA2006-09154,~~ XUGA PGID IT06PXIC296091PM 
and CPAN CSD2007-00042 (Spain), 
by grant FTC POCI/FP /81982 /2007  (Portugal),
by grant UCY-10.3.11.12 (Cyp\-rus), by INTAS grant
06-1000012-8861 and EU contract RII3-CT-2004-506078.

\end{document}